\documentclass{IEEEtran}
\usepackage{cite}
\usepackage{algorithm2e}
\usepackage{graphicx}
\usepackage{psfrag}
\usepackage{subfigure}
\usepackage{url}
\usepackage{amsmath}
\usepackage{array}
\usepackage{amssymb}
\usepackage{amsfonts}
\newtheorem{proposition}{Proposition}

\newtheorem{definition}{Definition}
\newtheorem{theorem}{Theorem}
\newtheorem{lemma}{Lemma}
\newtheorem{example}{Example}
\newtheorem{observation}{Observation}
\title{On Network-Error Correcting Convolutional Codes under the BSC Edge Error Model}
\author{
\authorblockN{K.~Prasad and B.~Sundar Rajan}
\authorblockA{Dept. of ECE, IISc, Bangalore 560012, India\\
Email: \{prasadk5,bsrajan\}@ece.iisc.ernet.in\\}
}
\date{\today}
\begin{document}
\maketitle
\thispagestyle{empty}
\begin{abstract}
Convolutional network-error correcting codes (CNECCs) are known to provide error correcting capability in acyclic instantaneous networks within the network coding paradigm under small field size conditions. In this work, we investigate the performance of CNECCs under the error model of the network where the edges are assumed to be statistically independent binary symmetric channels, each with the same probability of error $p_e$($0\leq p_e<0.5$). We obtain bounds on the performance of such CNECCs based on a modified generating function (the transfer function) of the CNECCs. For a given network, we derive a mathematical condition on how small $p_e$ should be so that only single edge network-errors need to be accounted for, thus reducing the complexity of evaluating the probability of error of any CNECC. Simulations indicate that convolutional codes are required to possess different properties to achieve good performance in low $p_e$ and high $p_e$ regimes. For the low $p_e$ regime, convolutional codes with good distance properties show good performance. For the high $p_e$ regime, convolutional codes that have a good \textit{slope} (the minimum normalized cycle weight) are seen to be good. We derive a lower bound on the slope of any rate $b/c$ convolutional code with a certain degree.
\end{abstract}
\section{Introduction}
\label{sec1}
Network coding as a means of increasing throughput in networks has been extensively studied in \cite{ACLY,KoM,CLY}. Block network-error correction for coherent network codes has been studied in \cite{YeC,Zha,YaY}. In all of these, the sufficient field size requirement for designing good block network-error correcting codes (BNECCs) is quite high. To be precise, the sufficient field size requirement for constructing a BNECC along with a network code which corrects network-errors due to any $t$ edges of the network being in error once in every $J$ network uses is such that 
$q > |{\cal{T}}|
\left(
\begin{array}{c}
J|\cal{E}| \\
2t
\end{array}
\right),
$ where ${\cal{T}}$ is the set of sinks. This requires every network-coding node of the network to perform multiplications of large degree polynomials over the base field each time it has to transmit, and therefore is computationally demanding. Moreover, the bound increases with the size of the network. It is therefore necessary to study network-error correcting codes which work under small field size conditions. 

Convolutional network-error correcting codes (CNECCs) were introduced in \cite{PrR} in the context of coherent network coding for acyclic instantaneous networks. The field size requirement for the CNECCs of \cite{PrR} is independent of the number of edges in the network and in general much smaller than what is demanded by BNECCs. Although the error correcting capability might not be comparable to that offered by BNECCs, the reduction in field size is a considerable advantage in terms of the computation to be performed at each coding node of the network. Also, the use of convolutional codes permits decoding using the Viterbi decoder, which is readily available. CNECCs with similar advantages for memory-free unit-delay acyclic networks were discussed in \cite{PrR2} and the benefit obtained in the performance of such CNECCs by using memory at the nodes of unit-delay networks was discussed in \cite{PrR3}. 

The CNECCs of \cite{PrR} were designed to correct network-errors which correspond to a set $\Phi$ of error patterns (subsets of the edge set) once in a certain number of network uses (a network use being the use of the edges of the network to transmit a number of symbols equal to the network code dimension). A similar error model (with $\Phi$ being all subsets of the edge set with $t$ edges) was considered in \cite{YeC,Zha,YaY}. While this error model allows code construction, it is less realistic because the errors corresponding to any error pattern in $\Phi$ are assumed to occur with equal probabilities. 

A more realistic error model would be to assume every edge $e$ in the network as a BSC with a certain cross-over probability ($p_e$) and with errors across different edges to be i.i.d. In this paper, we assume such an error model (with $p_e$ being the same for all edges) and analyze CNECCs over the binary field. Binary network codes together with this error model were studied in \cite{XiA}. The decoding of BNECCs under a similar probabilistic setting was discussed in \cite{BaL}. However, practical analysis and simulations of BNECCs under a probabilistic error setting is difficult because of the large field size demanded. On the other hand, the CNECCs developed in \cite{PrR} require small field sizes and thus facilitate analysis. The contributions and organization of this paper are as follows.
\begin{figure}[htbp]
\centering
\includegraphics[totalheight=2in,width=3in]{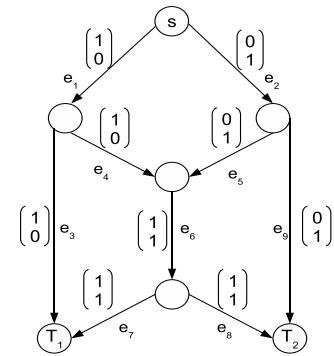}
\caption{Butterfly network}	
\label{fig:butterfly}	
\end{figure}
\begin{itemize}
\item After briefly discussing CNECCs for the network coding setup (Section \ref{sec2}), we present the error model for the network. If the edge cross-over probability $p_e << 0.5,$ then it is sufficient to compute only single edge network-error probabilities in the network thereby reducing the computations required to study the performance of CNECCs. For any network with a given number of edges, we derive a bound on how small this $p_e$ should be so that this assumption of ignoring multiple edge network-errors can be made safely. (Section \ref{sec3})

\item Expressions for the upper bound on the bit error probability of CNECCs are obtained based on a modified version of the augmented path generating function $\left(T(D,I)\right)$ of the CNECC being used. (Section \ref{sec4})

\item We analyze the performance of CNECCs on networks with a probabilistic error model using simulations with the butterfly network (Fig. \ref{fig:butterfly}) as an example. Simulations on the butterfly network indicate that different criteria apply for CNECCs to be good under low and high $p_e$ conditions. We therefore suggest different types of CNECCs under these two conditions. (Section \ref{sec5})

\item For high $p_e$ conditions, it is seen that those codes perform better which have a high value of slope, which is defined as follows.
\begin{definition}[\cite{JPZ}]
Given a minimal encoder of a rate $R=b/c$ convolutional code $\cal C$, the minimum normalized cycle weight
\begin{equation}
\label{slopedefinition}
\alpha:=\min_{o \in {\cal O}\backslash o_1}\left\{\frac{w_{_H}(o)}{l(o)}\right\}
\end{equation}
among all cycles $o \in {\cal O}$(the set of all cycles) in the state transition diagram of the encoder, except the zero cycle $o_1$ in the zero state, is called the slope $\alpha$ of the convolutional code $\cal C.$ Here $w_{_H}(o)$ indicates the Hamming weight accumulated by the output sequence while traversing the cycle $o$, and $l(o)$ is the length of the cycle in $c$-tuples.
\end{definition}

\item We derive a lower bound on the slope of any rate $b/c$ convolutional code over any finite field (Section \ref{sec6}),  and conclude with a short discussion of the paper and several directions for future research (Section \ref{sec7}). 
\end{itemize}
While CNECCs only over $\mathbb{F}_2$ are considered for the analyses and simulations of this paper, CNECCs over any field size can be studied using similar methods.
\section{Convolutional codes for network-error correction}
\label{sec2}
\subsection{Network model and network code}
An acyclic network can be represented as an acyclic directed multi-graph (a graph that can have parallel edges between nodes) ${\cal G}$ = ($\cal V,\cal E$) where $\cal V$ is the set of all vertices and $\cal E$ is the set of all edges in the network. Every edge in the directed multi-graph representing the network has unit \emph{capacity} (can carry utmost one symbol from $\mathbb{F}_2$).

Let $n$ be the mincut between the source $s$ and the set of sinks $\cal T$ and the dimension of the network code. An $n$-dimensional binary network code can be described by three matrices $A~\left(\text{of size}~n \times |{\cal E}|\right)$, $F~\left(\text{of size}~|{\cal E}| \times |{\cal E}|\right)$,and $B^T~\left(\text{of size}~|{\cal E}| \times n~ \text{for every sink}~ T\in {\cal T} \right)$, each having elements from $\mathbb{F}_2.$ Further details on the structure of these matrices can be found in \cite{KoM}. The network transfer matrix corresponding to a sink $T$ is an $n\times n$ binary matrix $M_T$ such that for any input $\boldsymbol{x}\in \mathbb{F}_2^n$, the output at sink $T\in \cal T$  is $\boldsymbol{x}M_T=\boldsymbol{x}AFB^T.$ 
\subsection{CNECCs}
For a given set of error patterns $\Phi$ and for some $k<n,$ a method of constructing rate $k/n$ convolutional codes was given in \cite{PrR} such that these CNECCs will correct network-errors which correspond to the patterns in $\Phi.$ For a given network with a network code, the definitions for the input and output convolutional code are as follows. 
\begin{definition}
An \textit{input convolutional code}, ${\cal C}_s$, corresponding to an acyclic network is a convolutional code of rate $~k/n (k < n)$ with a \textit{input generator matrix} $G_{I}(z)$ implemented at the source of the network.
\end{definition}
\begin{definition}
The \textit{output convolutional code} ${\cal C}_T$ corresponding to a sink node $T$ in the acyclic network is the $k/n$ convolutional code generated by the \textit{output generator matrix} $G_{O,{T}}(z)$ which is given by 
$
G_{O,{T}}(z) = G_I(z)M_{T},
$
with $M_T$ being the full rank network transfer matrix corresponding to an $n$-dimensional network code.
\end{definition}
It was shown in \cite{PrR} that errors corresponding to $\Phi$ can be corrected at all sinks as long as they are separated by a certain number of network uses. Moreover, a sink can achieve this error correcting capability by choosing to decode on either the input or the output convolutional codes depending upon their distance properties.  
\begin{example}
Table \ref{tab1} shows the network transfer matrices of the butterfly network of Fig. \ref{fig:butterfly} and an example of a CNECC along with the output convolutional codes at the two sinks.
\begin{table}[htbp]
\centering
\caption{Butterfly network of Fig. \ref{fig:butterfly} with the input convolutional code $G_I(z)=[1+z+z^2~~~1+z^2].$}
\begin{tabular}{|c|c|l|}
\hline
\textbf{Sink} & \textbf{Network transfer} & \textbf{Output convolutional code}\\
              & \textbf{matrix}           &                                   \\ 
\hline
$T_1$ & $M_{T_1}= \left(\begin{array}{cc}1 & 1 \\0  & 1 \end{array}\right)$ & $G_{O,T_1}(z)=[1+z+z^2~~~z]$\\
\hline
$T_2$ & $M_{T_2}=\left(\begin{array}{cc}1 & 0 \\1  & 1 \end{array} \right)$ & $G_{O,T_2}(z)=[z~~~1+z^2]$\\
\hline
\end{tabular}
\label{tab1}
\end{table}
\end{example}
\section{Network-errors in the BSC edge error model}
\label{sec3}
Any edge $e\in\cal E$ in the network is assumed to be a binary symmetric channel with probability of error being $p_e$ and errors on different edges are assumed to be i.i.d. A network-error is a vector $\boldsymbol{w}\in\mathbb{F}_2^{|{\cal E}|}$ with $1$s at those positions where the corresponding edge is in error. The probability of a network-error $\boldsymbol{w}\in\mathbb{F}_2^{|{\cal E}|}$ is then $p_e^{w_{_H}(\boldsymbol{w})}(1-p_e)^{|{\cal E}|-w_{_H}(\boldsymbol{w})}.$ 

Let $\boldsymbol{e}_{_T}$ denote the random error vector at sink $T.$ The probability that $\boldsymbol{e}_{_T}=\boldsymbol{y}\in\mathbb{F}_2^n$ is as follows.
\begin{align}
\label{eqn0}
p_{\boldsymbol{e}_{_T}}(\boldsymbol{y})&=\sum_{\boldsymbol{w}\in\mathbb{F}_2^{|{\cal E}|}:\boldsymbol{w}FB^T=\boldsymbol{y}}p_e^{w_{_H}(\boldsymbol{w})}(1-p_e)^{|{\cal E}|-w_{_H}(\boldsymbol{w})}\\
\nonumber
&=\sum_{i=1}^{{|\cal E|}}a_{i,\boldsymbol{y}}p_e^i(1-p_e)^{{|\cal E|}-i}
\end{align}
where $a_{i,\boldsymbol{y}}$ indicates the number of network-error vectors from $\mathbb{F}_2^{|\cal E|}$ with weight $i$, such that they result in the error vector $\boldsymbol{y}$ at sink $T.$

For any given network, it is essential to calculate the error probability of $\boldsymbol{e}_{_T}$ being any $\boldsymbol{y}\in\mathbb{F}_2^n$ for each sink $T\in \cal T$ in order to analyze the performance of any CNECC over the network. Equation (\ref{eqn0}) indicates that this involves a large number of computations even if the given network is small. However, if $p_e << 0.5,$ then it is sufficient to compute only single edge network-error probabilities for any particular error vector at any sink, thereby reducing the number of computations. In particular, suppose
\begin{align} 
\label{eqn8}
a_{1,\boldsymbol{y}}p_e(1-p_e)^{{|\cal E|}-1}  \geq \lambda \left(\sum_{i=2}^{{|\cal E|}}a_{i,\boldsymbol{y}}p_e^i(1-p_e)^{{|\cal E|}-i}\right)
\end{align}
for any error $\boldsymbol{y}$ at any sink $T$ with $a_{1,\boldsymbol{y}} \neq 0,$ for some $\lambda \geq 0.$ We then have the following upper bound. 
\begin{align}
\label{eqn9}
p_{\boldsymbol{e}_{_T}}(\boldsymbol{y})& \leq a_{1,\boldsymbol{y}}(1+\lambda^{-1})p_e(1-p_e)^{{|\cal E|}-1}~~~\forall~\boldsymbol{y} \in\mathbb{F}_2^n\backslash \{\boldsymbol{0}\} 
\end{align}
The probability of the error vector $\boldsymbol{e}_T$ being $\boldsymbol{0}\in\mathbb{F}_2^n$ is upper bounded independent of $\lambda$ as follows.
\begin{align} 
\label{eqn10}
p_{\boldsymbol{e}_{_T}}(\boldsymbol{0})& \leq 1-\sum_{\boldsymbol{y}\in\mathbb{F}_2^n\backslash \{\boldsymbol{0}\}:a_{1,\boldsymbol{y}}\neq 0}a_{1,\boldsymbol{y}}p_e(1-p_e)^{{|\cal E|}-1}
\end{align}

If $p_e$ is small enough so that (\ref{eqn8}) holds for some large $\lambda,$ then the upper bounds of (\ref{eqn9}) and (\ref{eqn10}) become tight, and hence single edge network-errors alone can be considered in the network without any significant loss of generality.
\subsection{An upper bound on $p_e$}
\label{subsec3a}
In this subsection, we obtain a sufficient upper bound on $p_e$ for a given network for (\ref{eqn8}) to hold so that only single edge network-error probabilities need to be calculated. This bound obtained holds for any network with a given number of edges and is independent of the network code chosen. It is seen that this bound on $p_e$ is inversely proportional to the number of edges in the network. This is a reasonable result because among the network-errors which result in some error vector at a sink, the difference between the number of multiple edge network-errors and the number of single edge network-errors would in general increase with the increase in network size, thus lowering the value of $p_e$ upto which (\ref{eqn8}) would hold. Towards calculating this bound, we first prove the following lemma.
\begin{lemma}
\label{lemma1}
For any integer $m\geq 1$ and $\forall~0 \leq p \leq 1,$ 
\[
(1-p)^m \geq 1-mp.
\]
\end{lemma}
\begin{IEEEproof}
For any $p > \frac{1}{m}, 1-mp < 0,$ and the proof is obvious. Therefore we prove the lemma only for $p \leq \frac{1}{m}.$ 

We have 
\begin{equation}
\label{eqn2}
(1-p)^m - 1 - mp = \sum_{i=2}^{m}
\left(
\begin{array}{c}
m\\
i
\end{array}
\right)
(-1)^ip^i
\end{equation}

Let $S$ be defined as 
\[
S:=\sum_{j=1}^{\lfloor\frac{m-1}{2}\rfloor}
\left(
\begin{array}{c}
m\\
2j
\end{array}
\right)
p^{2j} 
- 
\left(
\begin{array}{c}
m\\
2j+1
\end{array}
\right)
p^{2j+1}
\]

Therefore the R.H.S of (\ref{eqn2}) becomes 
\[
\sum_{i=2}^{m}
\left(
\begin{array}{c}
m\\
i
\end{array}
\right)
(-1)^ip^i = 
\left\{
\begin{array}{cc}
S & \text{if } m \text{ is odd }\\
S + p^m & \text{if } m \text{ is even }
\end{array}
\right.
\]
If $S \geq 0,$ the lemma is proved. Now every element inside the summation of $S$ is of the form
\begin{align}
\label{eqn3}
\left(
\begin{array}{c}
m\\
i
\end{array}
\right)
p^i
-
\left(
\begin{array}{c}
m\\
i+1
\end{array}
\right)
p^{i+1}
=
\left(
\begin{array}{c}
m\\
i
\end{array}
\right)
p^i
\left(
1-\frac{m-i}{i+1}p
\right)
\end{align}
If
$\left(
1-\frac{m-i}{i+1}p
\right)
\geq 0,
$  
then 
$
\left(
\begin{array}{c}
m\\
i
\end{array}
\right)
p^i
\left(
1-\frac{m-i}{i+1}p
\right)
\geq 0,$

Since $p \leq \frac{1}{m},$ we have
\[
\left(
1-\frac{m-i}{i+1}p
\right) \geq 
\left(
1-\frac{m-i}{i+1}.\frac{1}{m}
\right)
\geq 0.
\]
This means that every element in the summation of $S$ is non-negative, which means that $S \geq 0,$ hence proving the lemma.
\end{IEEEproof}
\begin{figure*}[tb]
\centering
\includegraphics[totalheight=4in,width=6.4in]{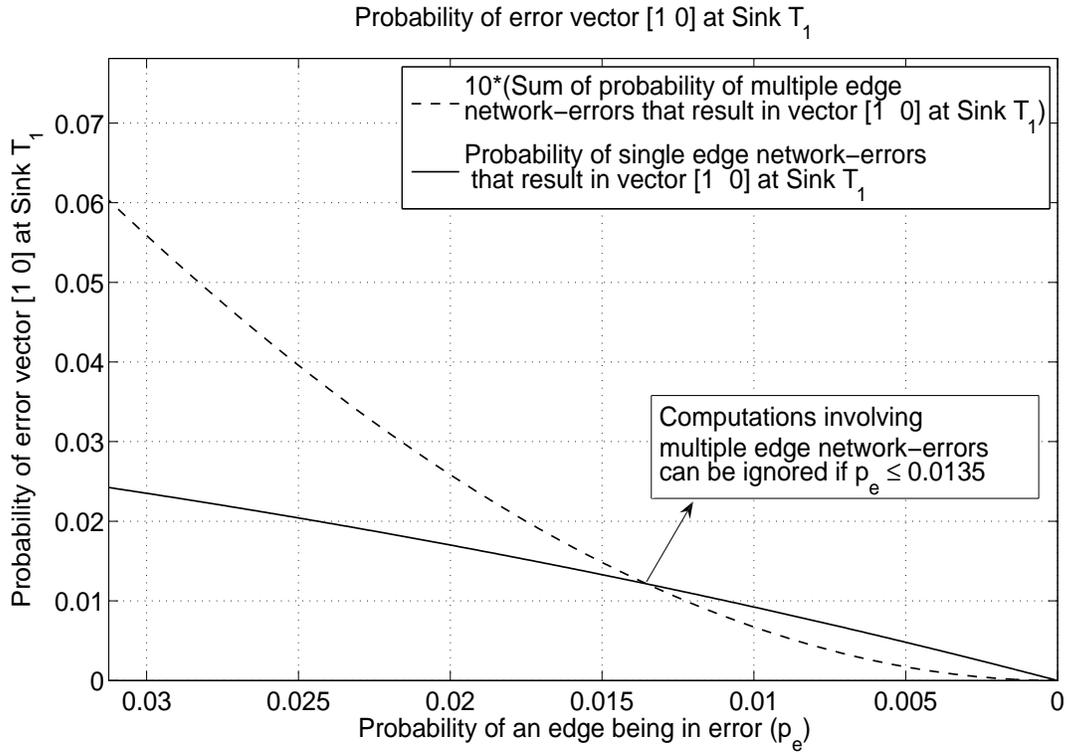}
\caption{Threshold $p_e$ for the butterfly network at Sink $T_1$}	
\label{fig:cutoffprob}	
\hrule
\end{figure*}
We now state and prove Proposition \ref{boundonpe} which gives the upper bound on $p_e$ for (\ref{eqn8}) to hold.
\begin{proposition}
\label{boundonpe}
For any error $\boldsymbol{y}$ at any sink $T$ with $a_{1,\boldsymbol{y}} \neq 0,$ the following holds
\[ 
a_{1,\boldsymbol{y}}p_e(1-p_e)^{{|\cal E|}-1}  \geq \lambda \left(\sum_{i=2}^{{|\cal E|}}a_{i,\boldsymbol{y}}p_e^i(1-p_e)^{{|\cal E|}-i}\right)
\]
if 
\[
p_e \leq \frac{1}{\left({|\cal E|}-1\right)\left(\lambda{|\cal E|}-\lambda+1\right)}.
\]
\end{proposition}
\begin{IEEEproof}
Since $a_{i,\boldsymbol{y}}\leq\left(\begin{array}{c}{|\cal E|}\\
i
\end{array}\right) \forall i$ and $a_{1,\boldsymbol{y}} \geq 1$ corresponding to such an error $\boldsymbol{y}$ as considered in the proposition, it is sufficient to consider the following case
\[
p_e(1-p_e)^{{|\cal E|}-1} \geq \lambda\left(\sum_{i=2}^{{|\cal E|}}
\left(
\begin{array}{c}{|\cal E|}\\
i
\end{array}
\right)
p_e^i(1-p_e)^{{|\cal E|}-i}\right)
\]
to get the bound on $p_e.$ Hence, we have
\begin{align}
\nonumber
&p_e(1-p_e)^{{|\cal E|}-1} \geq \lambda\left(\sum_{i=2}^{{|\cal E|}}
\left(
\begin{array}{c}
{|\cal E|}\\
i
\end{array}
\right)
p_e^i(1-p_e)^{{|\cal E|}-i} \right)\\ 
&p_e(1-p_e)^{{|\cal E|}-1} \geq \lambda\left(1-{|\cal E|}p_e(1-p_e)^{{|\cal E|}-1}-(1-p_e)^{|\cal E|}\right) \\
\label{eqn11}
&\Rightarrow (1-p_e)^{{|\cal E|}-1}\left( (\lambda{|\cal E|}+1)p_e+\lambda(1-p_e) \right) \geq \lambda
\end{align}
By Lemma \ref{lemma1}, the inequality of (\ref{eqn11}) holds if the following holds
\begin{align}
\nonumber
&\left(1-\left({|\cal E|}-1\right)p_e\right)\left( (\lambda{|\cal E|}+1)p_e+\lambda(1-p_e) \right) \geq \lambda \\
\label{eqn1}
&\Rightarrow \left(1-\left({|\cal E|}-1\right)p_e\right)\left(\lambda{|\cal E|}p_e-(\lambda-1)p_e+\lambda\right) \geq \lambda
\end{align}
Simplifying (\ref{eqn1}), we get
\begin{equation*}
p_e \leq \frac{1}{\left({|\cal E|}-1\right)\left(\lambda{|\cal E|}-\lambda+1\right)}
\end{equation*}
\end{IEEEproof}

The bound of Proposition \ref{boundonpe} holds for any network with $|{\cal E}|$ edges for a chosen $\lambda$ and in general is loose as indicated by Fig. \ref{fig:cutoffprob}. Having chosen $\lambda = 10,$ Fig. \ref{fig:cutoffprob} shows the single edge network-error probabilities and $10$ times the multiple edge network-error probabilities obtained using simulations with respect to varying $p_e,$ corresponding to the error vector $[1~~0]$ at Sink $T_1$ of the butterfly network. The threshold $p_e$ is approximately $0.0135,$ which is the lowest computed for any error vector at Sink $T_1.$ A similar value can be computed for Sink $T_2.$ This is approximately an order of magnitude greater than what the bound of Proposition \ref{boundonpe} indicates ($p_e\leq 0.00154$ for the butterfly network which has $9$ edges).
\begin{figure*}[tb]
\centering
\includegraphics[totalheight=4in,width=6.4in]{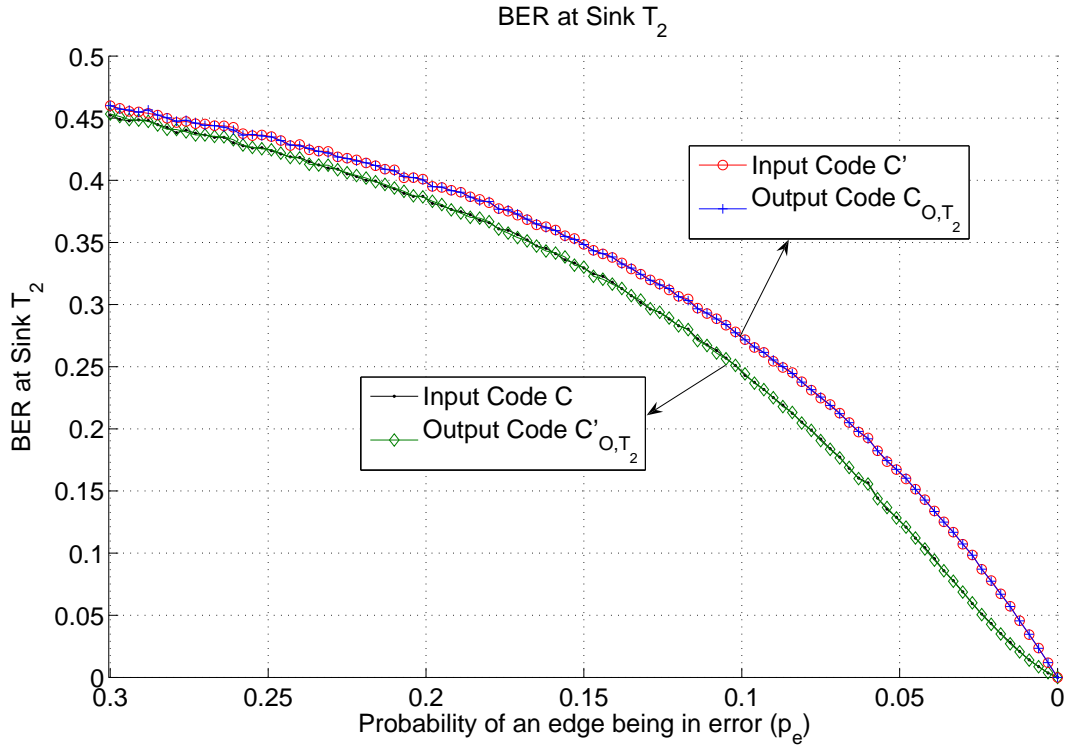}
\caption{BER at Sink $T_2$ for two CNECCs (Table \ref{tab3})}	
\label{fig:fordecoding}	
\hrule
\end{figure*}
\section{Bound on the bit error probability of a CNECC}
\label{sec4}
We can bound the bit error probability of a CNECC following \cite{ViO} upon a slight modification of its augmented generating function $T(D,I),$ which is a polynomial in $D$ and $I$  where any element of $T(D,I),$ say $bD^{d}I^{i},$ indicates $b$ number of paths which are unmerged with the all-zero codeword with a Hamming distance of $d$ and $i$ number of input 1s being encoded into the unmerged codeword segment. We compare the bound thus obtained with simulations on the butterfly network in Subsection \ref{subsec5b}.  

However, because the network coding channel has $\mathbb{F}_q^{n}$ inputs and $\mathbb{F}_q^{n}$ outputs, the generating function of the convolutional code needs to be modified to capture every $n$ bits transmitted at once. 

Therefore, we use the place-holders $D_{\boldsymbol{v}}$ for the branches of the state transition diagram with the output vector being $\boldsymbol{v}\in\mathbb{F}_q^n\backslash \{\boldsymbol{0}\}.$ The modified augmented generating function, $T(D_{000..01},...,D_{111..11},I)$, is thus the transfer function of the convolutional encoder with the state transition diagram with the branches weighted with appropriate $D_{\boldsymbol{v}}I^i.$

The bit error probability for a given rate $k/n$ CNECC for a sink $T$ is then bounded as 
\begin{equation}
\label{eqn5}
P_{b,T}\leq \left.\frac{1}{k}\frac{\partial T(D_{000..01},...,D_{111..11},I)}{\partial I}\right|_{I=1,D_{\boldsymbol{v}}=Z_{\boldsymbol{v},T}}  
\end{equation}
where 
\begin{equation*}
Z_{\boldsymbol{v},T} \equiv \sum_{\boldsymbol{y}\in\mathbb{F}_q^n}\sqrt{p_{e_{_T}}(\boldsymbol{y})p_{e_{_T}}(\boldsymbol{y}+\boldsymbol{v})}
\end{equation*}
is the Bhattacharyya bound on the pairwise error probability between $\boldsymbol{0}$ and $\boldsymbol{v},$ with $p_{e_{_T}}(\boldsymbol{y})$ being the probability that the error vector obtained at sink $T$ after applying the inverse of the network transfer matrix ($M_T$) is $\boldsymbol{y}.$ The partial derivative of (\ref{eqn5}) can be upper bounded  according to the numerical upper bound (\ref{eqn6}) shown at the top of the next page. 
\begin{figure*}
\begin{equation}
\label{eqn6}
\left.\frac{\partial T(D_{\boldsymbol{v}_1},...,D_{\boldsymbol{v}_{2^n-1}},I)}{\partial I}\right|_{I=1,D_{\boldsymbol{v}_i}=Z_{\boldsymbol{v}_i,T}}  < \frac{T(Z_{\boldsymbol{v}_1},...,Z_{\boldsymbol{v}_{2^n-1}},1+\epsilon)-T(Z_{\boldsymbol{v}_1},...,Z_{\boldsymbol{v}_{2^n-1}},1)}{\epsilon}~~~~~ 
\text{where}~~~\epsilon << 1. 
\end{equation}
\hrule
\end{figure*}
\begin{example}
\begin{figure}[htbp]
\centering
\includegraphics[totalheight=2.2in,width=3.2in]{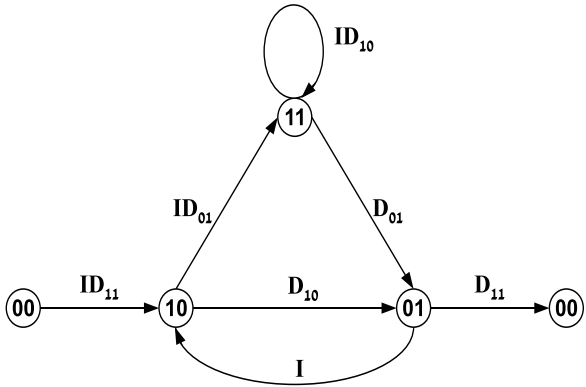}
\caption{State diagram of the code generated by $[1+z+z^2~~~1+z^2].$}	
\label{fig:statediag}	
\end{figure}
Fig. \ref{fig:statediag} shows the state transition diagram corresponding to a minimal encoder (controller canonical form) of the convolutional code generated by the matrix $[1+z+z^2~~~1+z^2].$	The modified augmented generating function can be obtained as
\begin{equation}
\label{eqn7}
T(D_{01},D_{10},D_{11},I)=\frac{I^2D_{11}^2\left(D_{01}^2-D_{10}^2\right)+ID_{11}^2D_{10}}{1+I^2\left(D_{10}^2-D_{01}^2\right)-2ID_{10}}.
\end{equation}
It can be noted that with $D_{\boldsymbol{v}}=D^{w_{_H}({\boldsymbol{v}})}$ in (\ref{eqn7}), the usual augmented generating function $T(D,I)$ of the code can be obtained. 
\end{example}
\section{Inference via simulation results}
\label{sec5}
\subsection{Decoding of CNECCs}
Given a $p_e$ value at which the network operates, any sink can choose to decode a CNECC either on the trellis of the input convolutional code or that of its output convolutional code, depending on their performance at the given $p_e$ value. Decoding on the output convolutional code is advantageous to any sink because it does not have to perform the network transfer matrix inversion before having to decode every time it receives the incoming symbols. 
\begin{table}[htbp]
\centering
\caption{$2$ CNECCs for the butterfly network (Fig. \ref{fig:butterfly}) with the output conv. codes at the Sink $T_2.$}
\begin{tabular}{|c|c|}
\hline
\textbf{Input convolutional code} & \textbf{Output convolutional code}\\
\textbf{generator matrix} & \textbf{generator matrix at Sink T2}\\
\hline
$[1+z~~~1]~~({\cal C})$ & $[z~~~1]~~({\cal C}_{O,T_2})$\\
\hline
$[1~~~z]~~({\cal C}')$ & $[1+z~~~z]~~({\cal C}_{O,T_2}')$\\
\hline
\end{tabular}
\label{tab3}
\end{table}
\begin{example}
Fig. \ref{fig:fordecoding} shows the performance of two CNECCs and their respective output convolutional codes (shown in Table \ref{tab3}) at sink $T_2$ of the butterfly network. It can be noted that for all $p_e$ values shown, code ${\cal C}_{O,T_2}'$ performs better than code ${\cal C}'.$ Thus if the code ${\cal C}'$ is used, sink $T_2$ can always decode on the trellis of ${\cal C}_{O,T_2}'.$ The opposite situation is observed for the pair ${\cal C}$ and ${\cal C}_{O,T_2}.$ It is therefore more beneficial for sink $T_2$ to decode on the trellis of ${\cal C}$ (after matrix inversion) for any $p_e \leq 0.25.$ For $p_e \geq 0.25,$ sink $T_2$ can decode on the trellis of ${\cal C}_{O,T_2}$, as the performance improvement obtained by decoding on ${\cal C}$ is negligible. 
\end{example}
\subsection{Coding for different values of $p_e$}
\label{subsec5b}
\begin{figure*}[tb]
\centering
\includegraphics[totalheight=4in,width=6.4in]{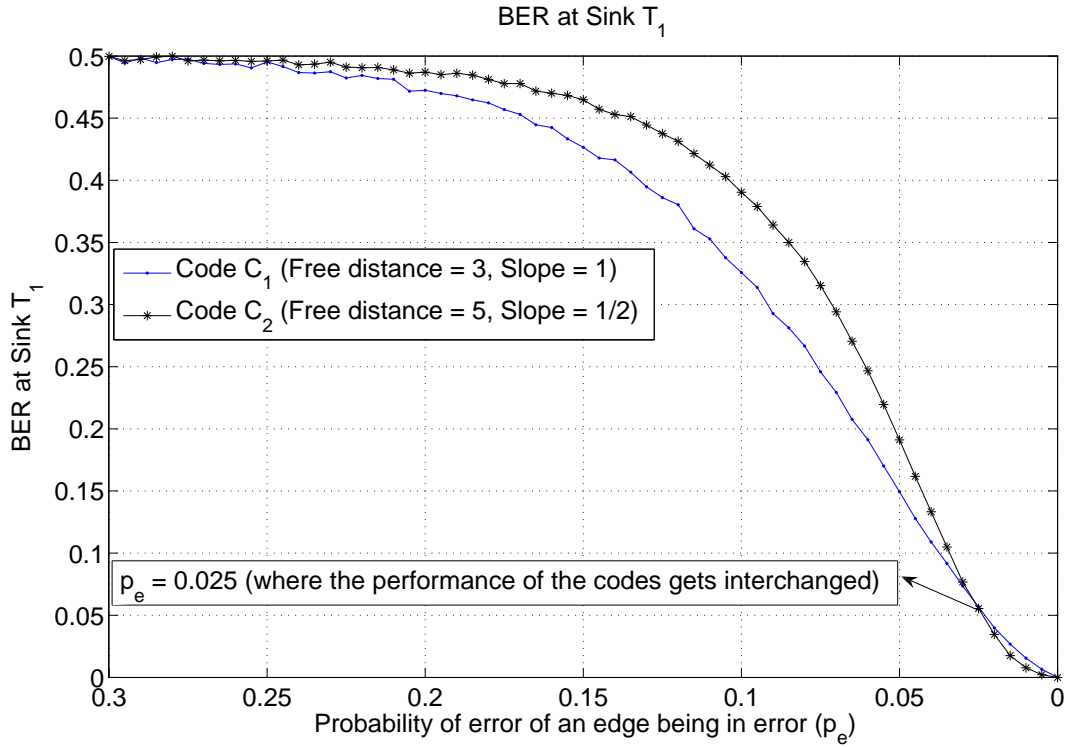}
\caption{BER and bounds on BER at Sink $T_1$ for $2$ codes}	
\label{fig:completecurve}	
\hrule
\end{figure*}
Fig. \ref{fig:completecurve} shows the performance of two different CNECCs (shown along with their properties in Table \ref{tab2}) at Sink $T_1$ of the butterfly network. Similar performances are seen at Sink $T_2.$  The decoding for all these CNECCs are done on the corresponding input convolutional code. 
\begin{table}[htbp]
\centering
\caption{CNECCs for the butterfly network}
\begin{tabular}{|c|c|c|}\hline
\textbf{CNECC generator matrix} & \textbf{Free distance} & \textbf{Slope}\\
\hline
$G_1(z)=[1+z~~1]~~~({\cal C}_1)$  & $3$ & $1$ \\
\hline
$G_1(z)=[1+z+z^2~~1+z^2]~~~({\cal C}_2)$ & $5$ & $1/2$\\
\hline
\end{tabular}
\label{tab2}
\end{table}
It is seen that there are two regimes of operation (for each pair of convolutional codes) where the performance of the codes get interchanged. This was already noticed in \cite{JPZ} in the context of AWGN channels. 
The value of $p_e$ for which these regimes becomes separated is not only dependent on the CNECC-pair chosen, but also on the network and the network code, and would probably decrease with the increase in the size of the network. 
\subsubsection{Coding for the low $p_e$ regime}
\begin{figure*}[tb]
\centering
\includegraphics[totalheight=4in,width=6.4in]{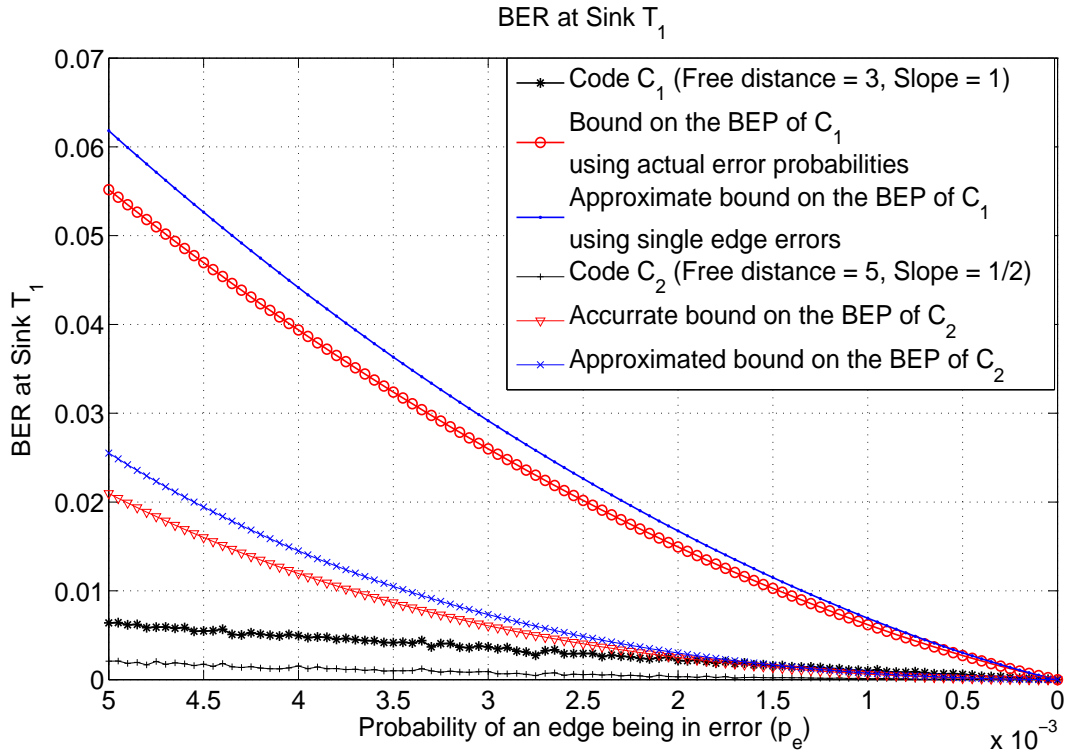}
\caption{BER and bounds on BER at Sink $T_1$ in the low $p_e$ regime}	
\label{fig:lowpecurve}	
\hrule
\end{figure*}
Fig. \ref{fig:lowpecurve} shows the performance of convolutional codes with different free distances on the butterfly network for low values of $p_e,$ along with the bounds on the bit-error probability evaluated according to Section \ref{sec4}. Codes with better distance spectra are good in the low $p_e$ regime. According to Fig. \ref{fig:completecurve}, this behavior is seen upto $p_e=0.025,$ however the bounds on the bit-error probability states become very loose beyond $p_e=0.005$ which is why the $p_e$ has been restricted to that value in Fig. \ref{fig:lowpecurve}.

Maximum Distance Separable (MDS) convolutional codes thus seem to be a good choice. The design of such convolutional codes along with the bounds on the field size requirement was discussed in \cite{PrR} for a fixed set of error patterns. If the value of $p_e$ is low enough, one might follow the design given in \cite{PrR} assuming the set of errors to be all possible single or double edge network-errors alone. 
\subsubsection{Coding for the high $p_e$ regime}
From Fig. \ref{fig:completecurve}, it is seen that codes with higher slopes are good for the high $p_e$ regime. The definition of the slope $\alpha$ of a convolutional code $\cal C$ is as in (\ref{slopedefinition}). For a given memory $m$ and free distance $d_{free}$, a convolutional code is said to be a \textit{maximum slope convolutional code} \cite{JPZ} if there exists no other code with a higher slope for the same memory and same free distance. Families of convolutional maximum slope convolutional codes were reported in \cite{JPZ}, discovered using computer search. 
\section{A lower bound on the slope of rate $b/c$ convolutional codes}
\label{sec6}
As seen in Subsection \ref{subsec5b}, codes with good slopes perform well in high $p_e$ conditions. It is therefore important to investigate the properties of the slope parameter and to come up with constructions which yield codes with good slopes. Upper bounds on the slope of convolutional codes were given in \cite{HuW,JPZ}. A lower bound on the slope of any rate $1/c$ convolutional code was given in \cite{HuW}. In this section, we derive a lower bound on the slope of any rate $b/c$ convolutional code over any finite field.

A primer on the basics of convolutional codes can be found in Appendix \ref{app1}. Towards obtaining a bound on the slope $\alpha$, we first give the following lemma. The proof of the following lemma is on the lines of Lemma 1 in \cite{PrR}.
\begin{lemma}
\label{deltazeroes}
Let $\cal C$ be a rate $b/c$ convolutional code with degree $\delta.$ For some $i\geq 0,$ if there exists a $\delta+1$ length partial codeword sequence
\[
\boldsymbol{v}_{[i,i+\delta]}: = \left[\boldsymbol{v}_i,\boldsymbol{v}_{i+1},. . . ,\boldsymbol{v}_{i+\delta}\right]
\]
where $\boldsymbol{v}_j=\boldsymbol{0} \in \mathbb{F}_q^c$ for $j=i,i+1,..i+\delta,$ then $\boldsymbol{v}_{[i,i+\delta]}$ has at least one cycle around the zero state of the corresponding minimal encoder of $\cal C.$ 
\end{lemma}
\begin{IEEEproof}
Let $G_{mb}(z)$ be a minimal basic generator matrix of $\left.{\cal C}\right).$ Let the ordered Forney indices (row degrees of $G_{mb}(z)$) be $\nu_1,\nu_2,. . . ,\nu_{b}=\nu_{max}$, and therefore $\delta$ being the sum of these indices. Then a systematic generator matrix($G_{sys}(z)$) for $\cal C$ that is equivalent to $G_{mb}(z)$ is of the form 
\begin{equation*}
G_{sys}(z)=T^{-1}(z)G_{mb}(z)
\end{equation*}
where $T(z)$ is a full rank $b \times b$ submatrix of $G_{mb}(z)$ with a delay-free determinant. We have the following observation. 
\begin{observation}
\label{degree}
The degree of $det\left(T(z)\right)$ is utmost $\delta.$ Also, we have the $(i,j)^{th}$ element $t_{i,j}(z)$ of $T^{-1}(z)$ as
\[
t_{i,j}(z)= \frac{Cofactor\left(T(z)_{j,i}\right)}{det\left(T(z)\right)}
\]
where $Cofactor(T(z)_{j,i}) \in \mathbb{F}_q[z]$ is the cofactor of the $(j,i)^{th}$ element of $T(z).$ The degree of $Cofactor(T(z)_{j,i})$ is utmost $\delta - \nu_j \leq \delta - \nu_1.$ 

Let $a_{i,j}(z) \in \mathbb{F}_q(z)$ represent the $(i,j)^{th}$ element of $G_{sys}(z),$ where 
\begin{eqnarray*}
a_{i,j}(z)=\sum_{k=1}^{b}t_{i,k}(z)g_{k,j}(z)~~~~~~~~ \\
~~~~~~~= \frac{\sum_{k=1}^{b}Cofactor(T(z)_{k,i})g_{k,j}(z)}{det\left(T(z)\right)}
\end{eqnarray*}
$g_{k,j}(z)$ being $(k,j)^{th}$ element of $G_{mb}(z).$ Therefore, the element $a_{i,j}(z)$ can be expressed as
\[
a_{i,j}(z) = \frac{p_{i,j}(z)}{det\left(T(z)\right)}
\]
where the degree of $p_{i,j}(z) \in \mathbb{F}_q[z]$ is utmost $\delta + \nu_{max} - \nu_1.$ Now if we divide $p_{i,j}(z)$ by $det\left(T(z)\right)$, we have
\begin{equation}
\label{eqn4}
a_{i,j}(z)= q_{i,j}(z) + \frac{r_{i,j}(z)}{det\left(T(z)\right)}
\end{equation}
where the degree of $q_{i,j}(z) \in \mathbb{F}_q[z]$ is utmost $\nu_{max}-\nu_1$, and the degree of $r_{i,j}(z)$ is utmost $\delta - 1.$ Because every element of $G_{sys}(z)$ can be reduced to the form in (\ref{eqn4}), we can have a realization of $G_{sys}(z)$ with utmost $\delta$ memory elements for each of the $b$ inputs. Let this encoder realization be known as $E_{sys}.$
\end{observation}
\begin{figure*}[tb]
\centering
\includegraphics[totalheight=3.5in,width=5in]{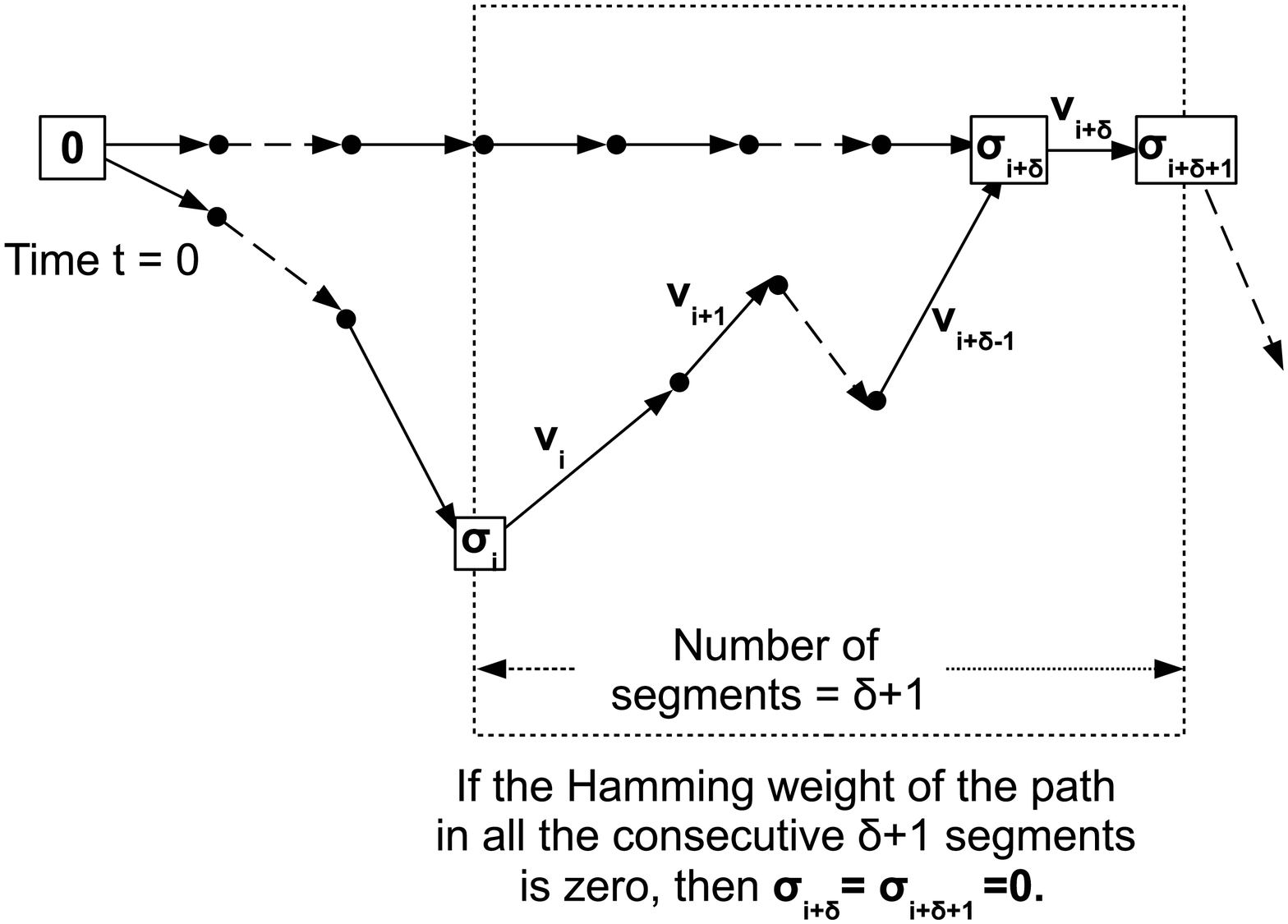}
\caption{The trellis corresponding to a systematic encoder of $\cal C$}	
\label{fig:alphabound}	
\hrule
\end{figure*}
Now we shall prove the lemma by contradiction. Let $v(z)$ be a codeword which contains the partial codeword sequence $v_{[i,i+\delta]}$ as follows: 
\[
\boldsymbol{v}(z)=\left[\boldsymbol{v}_0,\boldsymbol{v}_1,...,\boldsymbol{v}_i=\boldsymbol{0},\boldsymbol{0},...,\boldsymbol{0},\boldsymbol{v}_{i+\delta}=\boldsymbol{0},\boldsymbol{v}_{i+\delta+1},...\right]
\]

Let $\boldsymbol{u}_s(z)$ be the information sequence which when encoded into $\boldsymbol{v}(z)$ by the systematic encoder $E_{sys}.$ Because of the systematic property of $E_{sys}$, we must have that 
\[
\boldsymbol{u}_{s,i}=\boldsymbol{u}_{s,i+1}=...=\boldsymbol{u}_{s,i+\delta}=\boldsymbol{0}\in\mathbb{F}_q^b.
\]

By Observation \ref{degree}, $E_{sys}$ is an encoder which has utmost $\delta$ memory elements (for each input), and hence the state vector $\boldsymbol{\sigma}_{i+\delta}\in\mathbb{F}_q^{\delta}$ at time instant $i+\delta$ becomes zero as a result of $\delta$ zero input vectors. Fig. \ref{fig:alphabound} shows the scenario we consider. 

With another zero at time instant $i+\delta,$ there is a zero cycle. But we need to prove it for a minimal encoder, not a systematic one. So, we consider the codeword $\boldsymbol{v}(z),$ which can now be written as a unique sum of two code words $\boldsymbol{v}(z)=\boldsymbol{v}'(z)+\boldsymbol{v}''(z)$, where
\[
\boldsymbol{v}'(z)=\sum_{k=0}^{i+\delta}\boldsymbol{v}_kz^k=\left[\boldsymbol{v}_0,...,\boldsymbol{v}_{i}=\boldsymbol{0},...,\boldsymbol{v}_{i+\delta}=\boldsymbol{0},\boldsymbol{0},...\right]
\]
and 
\[
\boldsymbol{v}''(z)=\sum_{k=i+\delta+1}\boldsymbol{v}_kz^k=\left[\boldsymbol{0},\boldsymbol{0},...,\boldsymbol{0},\boldsymbol{0},\boldsymbol{v}_{i+\delta+1},...\right]
\]
where $\boldsymbol{0}\in \mathbb{F}_q^c$ and the uniqueness of the decomposition holds with respect to the positions of the zeros indicated in the two code words $\boldsymbol{v}'(z)$ and $\boldsymbol{v}''(z).$ 

Let $\boldsymbol{u}_{mb}(z)$ be the information sequence which is encoded into $\boldsymbol{v}(z)$ by a minimal realization $E_{mb}$ of a minimal basic generator matrix $G_{mb}(z)$ (a minimal encoder). Then we have
\[
\boldsymbol{u}_{mb}(z)=\boldsymbol{u}'_{mb}(z)+\boldsymbol{u}''_{mb}(z)
\]
where $\boldsymbol{u}'_{mb}(z)$ and $\boldsymbol{u}''_{mb}(z)$ are encoded by $E_{mb}$ into $\boldsymbol{v}'(z)$ and $\boldsymbol{v}''(z)$ respectively.

By the \textit{predictable degree property} (PDP) \cite{JoZ} of minimal basic generator matrices, we have that for any polynomial code sequence $\boldsymbol{v}(z)$,
\[
deg\left(\boldsymbol{v}(z)\right)=\max_{1\leq l \leq b}\left\{deg\left(\boldsymbol{u}_{mb,l}(z)\right)+\nu_l\right\}.
\]
where $\boldsymbol{u}_{mb,l}(z) \in \mathbb{F}_q[z]$ represents the information sequence corresponding to the $l^{th}$ input, and $deg$ indicates the degree of the polynomial. Therefore, by the PDP property, we have that $deg\left(\boldsymbol{u}'_{mb}(z)\right) < i$, since  $deg\left(\boldsymbol{v}'(z)\right)<i$. 

Also, it is known that in the trellis of corresponding to a minimal realization of a minimal-basic generator matrix, there exists no non-trivial transition from the all-zero state to a non-zero state that produces a zero output. Therefore we have $deg\left(\boldsymbol{u}''_{mb}(z)\right) \geq i+\delta+1$, with equality being satisfied if $\boldsymbol{v}_{i+\delta+1}\neq \boldsymbol{0}.$ Therefore, $u_{mb}(z)$ is of the form
\begin{eqnarray*}
\boldsymbol{u}_{mb}(z)=\boldsymbol{u}'_{mb}(z)+\boldsymbol{u}''_{mb}(z) ~~~~~~~~~~~~~~~~~~~~~~~~~~\\
=\sum_{k=1}^{i-1}\boldsymbol{u}'_{mb,k}z^k + \sum_{k=i+\delta+1}^{\infty}\boldsymbol{u}''_{mb,k}z^k ~~~~~~~~~~~~\\
\boldsymbol{u}_{mb}(z)=\left[\boldsymbol{u}'_{mb,0},..,\boldsymbol{u}'_{mb,i-1},\boldsymbol{0},\boldsymbol{0},..\right]~~~~~~~~~~~~~~~~~~~~~~~~~\\
~~~~~~~~~~~~~~~~~+\left[\boldsymbol{0},..,\boldsymbol{0},\boldsymbol{u}''_{mb,i+\delta+1},\boldsymbol{u}''_{mb,i+\delta+2},..\right]~~~~~~~~~~~~~~
\end{eqnarray*}
i.e, if
\[
\boldsymbol{u}_{mb}(z)=\left[\boldsymbol{u}_{mb,0},\boldsymbol{u}_{mb,1},...,\boldsymbol{u}_{mb,i},...,\boldsymbol{u}_{mb,i+\delta},\boldsymbol{u}_{mb,i+\delta+1},..\right]
\]
then $\boldsymbol{u}_{mb,i}=\boldsymbol{u}_{mb,i+1}=...=\boldsymbol{u}_{mb,i+\delta}=\boldsymbol{0} \in \mathbb{F}_q^b.$ 

With the minimal encoder $E_{mb}$ which has $\nu_{max}$ memory elements, these $\delta+1$ consecutive zeros of $\boldsymbol{u}_{mb}(z)$ would result in the state vector $\boldsymbol{\sigma}_{mb,t}$ becoming zero for all time instants from $i+\nu_{max}$ to $i+\delta+1,$ i.e.,
\[
\boldsymbol{\sigma}_{mb,i+\nu_{max}}=\boldsymbol{\sigma}_{mb,i+\nu_{max}+1}=...=\boldsymbol{\sigma}_{mb,i+\delta+1}=\boldsymbol{0}\in\mathbb{F}_q^{\nu_{max}}.
\]
With $\nu_{max} \leq \delta,$ the path traced by $\boldsymbol{v}_{[i,i+\delta]}$ traces at least one zero cycle on the trellis corresponding to the minimal encoder. This concludes the proof. 
\end{IEEEproof}
We shall now prove the bound on $\alpha.$
\begin{theorem}
The slope $\alpha$ of a rate $b/c$ convolutional code $\cal C$ with degree $\delta$ is lower bounded as
\[
\alpha \geq \frac{1}{\delta+1}. 
\]
\end{theorem}
\begin{IEEEproof}
First we note the fact that every path in the state transition diagram is either a cycle or a part of a cycle. By Lemma \ref{deltazeroes}, the path traced by any partial codeword sequence with $\delta+1$ consecutive zero components in the state transition diagram of the minimal encoder would have a cycled around in the zero state at least once. The definition of $\alpha$ excludes a cycle around the zero state, and therefore paths (partial codeword sequences) which have $\delta+1$ consecutive zero components cannot be considered to measure $\alpha$ since they would ultimately result in a zero cycle. However, Lemma \ref{deltazeroes} also implies that any path in the state transition diagram which does not include the zero cycle must therefore accumulate at least 1 Hamming weight in every $\delta+1$ transitions. Thus we have proved that the slope is lower bounded as:
\[
\alpha \geq \frac{1}{\delta+1}.  
\]
\end{IEEEproof}
\section{Discussion}
\label{sec7}
The performance of CNECCs under the BSC edge error model has been analyzed using theoretical bounds and simulations. A sufficient upper bound on the edge cross-over probability $p_e$ has been obtained, so that if $p_e$ is below this bound, the complexity of analysis can be reduced greatly by considering only single edge network-errors. Codes with better distance spectra and those with good slopes are seen to perform well under different conditions on the cross-over probability. A lower bound on the slope of any convolutional code is also obtained. Several interesting problems remain in this context including the following. 
\begin{itemize}
\item Studying the soft-decision decoding performance of CNECCs. 
\item Constructions of convolutional codes with good slopes. 
\item In large networks, error probabilities at the sinks could be large even for negligible $p_e$ values. It would be interesting to look at the existing network error correction schemes for such networks, and compare them with schemes which involve coding over smaller subnetworks.
\end{itemize}
\section*{Acknowledgment} This work was supported  partly by the DRDO-IISc program on Advanced Research in Mathematical Engineering to B.~S.~Rajan.

\appendices%
\section{Convolutional codes-Basic Results}
\label{app1}
We review the basic concepts related to convolutional codes, used extensively throughout the rest of the paper. For $q,$ power of a prime, let $\mathbb{F}_q$ denote the finite field with $q$ elements,  $\mathbb{F}_q[z]$ denote \textit{the ring of univariate polynomials} in $z$ with coefficients from $\mathbb{F}_q,$  $\mathbb{F}_q(z)$ denote \textit{the field of rational functions} with variable $z$ and coefficients from $\mathbb{F}_q$ and $\mathbb{F}_q[[z]]$ denote  \textit{the ring of formal power series} with coefficients from $\mathbb{F}_q$. Every element of $\mathbb{F}_q[[z]]$ of the form $x(z)=\sum_{i=0}^\infty x_iz^i, x_i \in \mathbb{F}_q$. Thus, $\mathbb{F}_q[z] \subset \mathbb{F}_q[[z]]$. We denote the set of $n$-tuples over $\mathbb{F}_q[[z]]$ as $\mathbb{F}_q^n[[z]]$. Also, a rational function $x(z)=\frac{a(z)}{b(z)}$ with $b(0) \neq 0$ is said to be \textit{realizable}. A matrix populated entirely with realizable functions is called a realizable matrix.

For a convolutional code, the \textit{information sequence} $\boldsymbol{u} = \left[\boldsymbol{u}_0,\boldsymbol{u}_1,...,\boldsymbol{u}_t\right](\boldsymbol{u}_i\in\mathbb{F}_q^b)$ and the \textit{codeword sequence} (output sequence) $\boldsymbol{v} = \left[\boldsymbol{v}_0,\boldsymbol{v}_1,...,\boldsymbol{v}_t\right]\left(\boldsymbol{v}_i\in\mathbb{F}_q^c\right)$ can be represented in terms of the delay parameter $z$ as 	
\begin{eqnarray*}
\boldsymbol{u}(z)=\sum_{i=0}^t \boldsymbol{u}_i z^i ~~~ \mbox{  and  }~~~
\boldsymbol{v}(z)=\sum_{i=0}^t \boldsymbol{v}_i z^i
\end{eqnarray*}
\begin{definition}[\cite{JoZ}]
A \textit{convolutional code}, ${\cal C}$ of rate $~b/c~(b~<~c)$ is defined as 
\[
{\cal C} = \{ \boldsymbol{v}(z)\in\mathbb{F}_q^{c}[[z]]~|~ \boldsymbol{v}(z)=\boldsymbol{u}(z)G(z) \}
\] 
where $G(z)$ is a $b \times c$  \textit{generator matrix} with entries from $\mathbb{F}_q(z)$ and rank $b$ over $\mathbb{F}_q(z)$, and $\boldsymbol{v}(z)$ being the codeword sequence arising from the information sequence, $\boldsymbol{u}(z)\in\mathbb{F}_q^{b}[[z]]$.
\end{definition}

Two generator matrices are said to be \textit{equivalent} if they encode the same convolutional code. A \textit{polynomial generator matrix}\cite{JoZ} for a convolutional code $\cal C$ is a generator matrix for $\cal C$ with all its entries from $\mathbb{F}_q[z]$. It is known that every convolutional code has a polynomial generator matrix \cite{JoZ}. Also, a generator matrix for a convolutional code is \textit{catastrophic}\cite{JoZ} if there exists an information sequence with infinitely many non-zero components, that results in a codeword with only finitely many non-zero components.

For a polynomial generator matrix $G(z)$, let $g_{ij}(z)$ be the element of $G(z)$ in the $i^{th}$ row and the $j^{th}$ column, and 
\[
\nu_i:=\max_{j} deg(g_{ij}(z))
\]
be the $i^{th}$ \textit{row degree} of $G(z)$. Let 
\[
\delta: = \sum_{i=1}^{b}\nu_i
\]
be the \textit{degree} of $G(z).$
\begin{definition}[\cite{JoZ} ]
A polynomial generator matrix is called \textit{basic} if it has a polynomial right inverse. It is called \textit{minimal} if its degree $\delta$ is minimum among all generator matrices of $\cal C$.
\end{definition}

Forney in \cite{For} showed that the ordered set $\left\{\nu_{1},\nu_{2},...,\nu_{b}\right\}$ of row degrees (indices) is the same for all minimal basic generator matrices of $\cal C$ (which are all equivalent to one another). Therefore the ordered row degrees and the degree $\delta$ can be defined for a convolutional code $\cal C.$ Also, any minimal basic generator matrix for a convolutional code is non-catastrophic. 

\begin{definition}[\cite{JoZ} ]
A \textit{convolutional encoder} is a physical realization of a generator matrix by a linear sequential circuit. Two encoders are said to be \textit{equivalent encoders} if they encode the same code. A \textit{minimal encoder} is an encoder with the minimal number of memory elements among all equivalent encoders.
\end{definition}

Given an encoder with $\delta '$ memory elements for the code $\cal C$, we can associate a vector $\boldsymbol{\sigma}_t \in \mathbb{F}_q^{\delta '}$ whose components indicate the states of the $\delta '$ memory elements at time instant $t.$  

The weight of a vector $\boldsymbol{v}(z) \in \mathbb{F}_q^{c}[[z]]$ is the sum of the Hamming weights (over $\mathbb{F}_q$) of all its $\mathbb{F}_q^{c}$-coefficients. Then we have the following definitions.
\begin{definition}[\cite{JoZ}]
The \textit{free distance} of a convolutional code $\cal C$ is given as 
\[
d_{free}({\cal C})=min\left\{wt(\boldsymbol{v}(z))|\boldsymbol{v}(z)\in{\cal C},\boldsymbol{v}(z)\neq 0\right\}
\]
\end{definition}

\begin{thebibliography}{160}
\bibitem{ACLY}
R. Ahlswede, N. Cai, R. Li and R. Yeung, ``Network Information Flow'', IEEE Transactions on Information Theory, vol.46, no.4, July 2000, pp. 1204-1216.

\bibitem{CLY}
N. Cai, R. Li and R. Yeung, ``Linear Network Coding'', IEEE Transactions on Information Theory, vol. 49, no. 2, Feb. 2003, pp. 371-381.

\bibitem{KoM}
R. Koetter and M. Medard, ``An Algebraic Approach to Network Coding'', IEEE/ACM Transactions on Networking, vol. 11, no. 5, Oct. 2003, pp. 782-795. 

\bibitem{YeC}
Raymond W. Yeung and Ning Cai, ``Network error correction, part 1 and part 2'', Comm. in Inform. and Systems, vol. 6, 2006, pp. 19-36.

\bibitem{Zha}
Zhen Zhang, ``Linear network-error Correction Codes in Packet Networks'', IEEE Transactions on Information Theory, vol. 54, no. 1, Jan. 2008, pp. 209-218. 

\bibitem{YaY}
Shenghao Yang and Yeung, R.W., ``Refined Coding Bounds for network error Correction'', ITW on Information Theory for Wireless Networks, July 1\text{-}6, 2007, Bergen, Norway, pp. 1-5.

\bibitem{PrR}
K. Prasad and B. Sundar Rajan, ``Convolutional codes for Network-error correction'', arXiv:0902.4177v3 [cs.IT], August 2009, available at: http://arxiv.org/abs/0902.4177. A shortened version of this paper is to appear in the proceedings of Globecom 2009, Nov. 30 - Dec. 4, Honolulu, Hawaii, USA.

\bibitem{PrR2}
K. Prasad and B. Sundar Rajan, ``Network error correction for unit-delay, memory-free networks using convolutional codes'', arXiv:0903.1967v3 [cs.IT], Sep. 2009, available at: http://arxiv.org/abs/0903.1967. A shortened version of this paper is to appear in the proceedings of ICC 2010, May 23 - 27, Capetown, South Africa.

\bibitem{PrR3}
K. Prasad and B. Sundar Rajan, ``Single-generation network coding for networks with delay'', arXiv:0909.1638v1 [cs.IT], Sep. 2009, available at: http://arxiv.org/abs/0909.1638. A shortened version of this paper is to appear in the proceedings of ICC 2010, May 23 - 27, Capetown, South Africa.

\bibitem{XiA}
Ming Xiao and Aulin, T.M., ``A Physical Layer Aspect of Network Coding with Statistically Independent Noisy Channels'', Proceedings of ICC 2006, June 1-15, Istanbul, Turkey, pp. 3996-4001.

\bibitem{BaL}
Bahramgiri, H. and Lahouti, F., ``Block network error control codes and syndrome-based maximum likelihood decoding'', Proceedings of ISIT 2008, July 6-11, Toronto, Canada, pp. 807-811.

\bibitem{JPZ}
R. Jordan, V. Pavlushkov, and V.V. Zyablov, ``Maximum Slope Convolutional Codes'', IEEE transactions of information theory, Vol. 50, No. 10, Oct. 2004, pp. 2511-2522

\bibitem{ViO}
Andrew J. Viterbi, James K.Omura, ``Principles of Digital Communication and Coding'', McGraw-Hill, 1979.
\bibitem{JoZ}
R. Johannesson and K.S Zigangirov, ``Fundamentals of Convolutional Coding'', John Wiley, 1999.

\bibitem{HuW}
G. K. Huth and C. L. Weber, ``Minimum weight convolutional codewords of finite length'', IEEE Trans. Inform. Theory, Vol. IT-22, Mar. 1976, pp. 243-246.

\bibitem{For}
G. D. Forney, ``Minimal bases of Rational Vector Spaces with applications to multivariable linear systems'', SIAM J. Contr., vol. 13, no. 3, 1975, pp. 493-520.
\end{thebibliography}
\end{document}